\documentstyle[aps,epsf,eqsecnum]{revtex}
\begin{document}
\preprint{UAHEP 979}
\draft
\title{Perturbations in the Kerr-Newman Dilatonic
Black Hole Background: \\
Maxwell Waves, the Dilaton Background and Gravitational Lensing}
\author{R. Casadio}
\address{Dipartimento di Fisica, Universit\`a di Bologna, and \\
I.N.F.N., Sezione di Bologna, \\
via Irnerio 46, 40126 Bologna, Italy}
\author{B. Harms}
\address{Department of Physics and Astronomy,
The University of Alabama\\
Box 870324, Tuscaloosa, AL 35487-0324}
\maketitle
\begin{abstract}
In this paper we continue the analysis of one of our previous papers
and study the
affect of the existence of a non-trivial dilaton background on the
propagation of electromagnetic waves in the Kerr-Newman dilatonic
black hole space-time.
For this purpose we again employ the double expansion in both the
background electric charge and the wave parameters of the relevant
quantities in the Newman-Penrose formalism and then identify the
first order at which the dilaton background enters Maxwell equations.
We then assume that gravitational and dilatonic waves are negligible
(at that order in the charge parameter) with respect to electromagnetic
waves and argue that this condition is consistent with the solutions
already found in the previous paper.
Explicit expressions are given for the asymptotic behaviour of scattered
waves, and a simple physical model is proposed in order to test the
effects.
An expression for the relative intensity is obtained for
Reissner-Nordstr\"om dilaton black holes using geometrical optics.
A comparison with the approximation of geometrical optics for
Kerr-Newman dilaton black holes shows that at the order to which
the calculations are carried out gravitational lensing of optical
images cannot probe the dilaton background.
\end{abstract}
\pacs{4.60.+n, 11.17.+y, 97.60.Lf}
\section{Introduction}
\label{intro}
In Ref.~\cite{chlc} we started from the low energy effective action
describing the Einstein-Maxwell theory interacting with a dilaton
of arbitrary coupling constant $a$ in four dimensions
(we always set $c=G=1$ unless differently stated),
\begin{eqnarray}
S = {1\over16\,\pi}\,\int d^4x\,\sqrt{-g}\,\left[R-{1\over2}\,
(\nabla\phi)^2-e^{-a\,\phi}\,F^2\right]
\ ,
\label{action}
\end{eqnarray}
and, on expanding the fields in terms of the charge-to-mass ratio
$Q/M$ of the source, we obtained the static solution of the field
equations corresponding to a Kerr-Newman dilatonic (KND) black hole
rotating with arbitrary angular momentum.
Our solution reduces to the Kerr-Newman (KN) metric
(see {\em e.g.} \cite{chandra})
for $a=0$ and to the Kaluza-Klein solution \cite{kk} for $a=\sqrt{3}$,
but differs from the exact Reissner-Nordstr\"om dilatonic (RND) black
hole \cite{rnd} in the zero angular momentum limit \cite{hh}.
In Ref.~\cite{kndw}, to which we refer for all the definitions
and notation, we used these solutions to write the wave equations
for the various field modes.
Our method is to double expand each field in powers of the electric
charge and the wave parameter.
Substituting these expansions into Maxwell's equations, the dilaton
equation and Einstein's equations then give (inhomogeneous) wave
equations for the coefficients of the expansion to any desired
accuracy.
\par
We have already given explicit expressions for Maxwell's equations for
the coefficients linear in the wave parameter for the three lowest
orders of the charge parameter and have found the asymptotic form of the
solutions linear in the wave and charge parameters (order $(1,1)$ in the
notation of \cite{kndw}).
The latter correspond to Maxwell waves produced by dilaton waves
scattered by the static electromagnetic background.
\par
In this paper we analyze the Maxwell waves produced by the scattering
of electromagnetic waves by the static dilaton background which emerge at
order $(1,2)$.
Since Maxwell's equations at order $(1,2)$ contain waves of all kinds
\cite{rn}, finding analytic solutions seems an intractable problem.
We thus pursue a suggestion already introduced in \cite{kndw} and
assume that the dilaton waves and the gravitational waves are negligible
with respect to Maxwell waves which allows us to greatly simplify
the equations for electromagnetic waves.
In section~\ref{check} we check the consistency of such a working ansatz
and argue that it is valid at order (1,2) when one considers the
solutions
at order $(1,1)$ which we have found in \cite{kndw}.
In section~\ref{sol} we compute the asymptotic behaviour at large
distance
from the hole of both outgoing and ingoing modes of the reduced Maxwell
equations and propose a physical onset to detect the dilaton background
in a binary system from the scattered pattern of radiation emitted by
the star companion.
In section~\ref{def} we show an alternative way of computing the variation
of the flux of a null wave scattered by RND and KND black holes in the
approximation of geometrical optics.
This approach reflects the qualitatively different natures of the two
kinds of metrics and proves that light paths are not affected be
the dilaton background in KND at the computed order in the charge-to-mass
expansion.
We also compare this result with the wave approach of the previous
sections.
\section{Wave equations at order (1,2)}
\label{check}
We employ the double expansion of perturbations \cite{kndw}
of the static solution given in Ref.~\cite{chlc} in order to write
the wave equations for the various field modes,
\begin{eqnarray}
&&\phi(t,r,\theta,\varphi)=\sum\limits_{p,n}\,g^p\,Q^n\,\phi^{(p,n)}
\nonumber \\
&&\phi_i(t,r,\theta,\varphi)=\sum\limits_{p,n}\,g^p\,Q^n\,
\phi^{(p,n)}_i\ ,
\ \ \ i=0,1,2
\nonumber \\
&&G(t,r,\theta,\varphi)=\sum\limits_{p,n}\,g^p\,Q^n\,G^{(p,n)}
\ ,
\label{(l,n)}
\end{eqnarray}
where $g$ is the same wave parameter for the dilaton ($\phi$), Maxwell
($\phi_i$) and gravitational (collectively denoted by $G$) field
quantities.
Of course, we will only study the linear ($p=1$) case and set $g=1$ from
now on.
We assume the linear perturbations have the following time and azimuthal
dependence
\begin{eqnarray}
\phi^{(1,n)}(t,r,\theta,\varphi)&=&k_d\,
e^{i\,\bar\omega\,t+i\,m\,\varphi}\,\phi^{(1,n)}(r,\theta)
\nonumber \\
\phi_i^{(1,n)}(t,r,\theta,\varphi)&=&k_{EM}\,
e^{i\,\bar\omega\,t+i\,m\,\varphi}\,
\phi_i^{(1,n)}(r,\theta)\ ,\ \ \ i=0,1,2
\nonumber \\
G^{(1,n)}(t,r,\theta,\varphi)&=&k_G\,
e^{i\,\bar\omega\,t+i\,m\,\varphi}\,
G^{(1,n)}(r,\theta)
\ ,
\end{eqnarray}
where $k_d$, $k_{EM}$ and $k_G$ are parameters.
We recall here that each function of $r$ and $\theta$ on the R.H.S.s
above implicitly carries an extra integer index, $m$, anda continuous
dependence on the frequency $\bar\omega$.
\par
As shown in \cite{kndw}, it is at order $(1,2)$ that the effect
induced by the dilaton background appears in all of the equations.
Since at this order the different kinds of waves do not disentangle
even for $a=0$ \cite{rn},
we need a working ansatz to obtain a manageable set of equations.
As already suggested in \cite{kndw}, we shall tentatively assume
\begin{eqnarray}
k_{EM}\gg k_d,k_G
\ ,
\label{kkk}
\end{eqnarray}
and neglect both dilaton and gravitational waves with respect to Maxwell
waves.
This hypothesis is equivalent to assuming the existence of (space-time)
boundary conditions such that the gravitational and dilaton contents
of the wave field are negligibly small when compared to the
electromagnetic
sources, a condition that is not automatically consistent and will
be checked in the following for the whole set of field equations.
\par
For completeness, we observe that, in case $k_{EM}\sim k_d,k_G$ the
leading contributions to the electromagnetic waves come from order (1,1)
and have already been found in \cite{kndw}.
Therefore the present notes together with \cite{kndw} should cover most
of the leading order electromagnetic physics that one can extract from
the study of classical waves on the KND background.
\subsection{Gravitational equations}
\label{G_w}
In \cite{kndw} we considered the gravitational field to
be determined by the following three non vacuum equations
in the Newman-Penrose (NP) formalism,
\begin{eqnarray}
(\hat\delta^*-4\,\tilde\alpha+\pi)\,\Psi_0
-(\hat D-4\,\tilde\rho-2\,\epsilon)\,\Psi_1
-3\,\kappa\,\Psi_2
=(\hat\delta+\pi^*-2\,\tilde\alpha^*-2\,\beta)\,R_{11}
\nonumber \\
-(\hat D-2\,\epsilon-2\,\tilde\rho^*)\,R_{12}
+2\,\sigma\,\,R_{21}
-2\,\kappa\,\,R_{22}
-\kappa^*\,\,R_{13}
\nonumber \\
\nonumber \\
(\hat\Delta-4\,\gamma+\mu)\,\Psi_0
-(\hat\delta-4\,\tau-2\,\beta)\,\Psi_1
-3\,\sigma\,\Psi_2
=(\hat\delta+2\,\pi^*-2\,\beta)\,R_{12}
\nonumber \\
-(\hat D-2\,\epsilon+2\,\epsilon^*-\tilde\rho^*)\,R_{13}
-\lambda^*\,\,R_{11}
-2\,\sigma\,\,R_{22}
-2\,\kappa\,\,R_{23}
\nonumber \\
\nonumber \\
(\hat D-\tilde\rho-\tilde\rho^*-3\,\epsilon+\epsilon^*)\,\sigma
-(\hat\delta-\tau+\pi^*-\tilde\alpha^*-3\,\beta)\,\kappa
-\Psi_0=0
\ .
\label{grav}
\end{eqnarray}
The Ricci tensor terms are given by the Einstein field equations,
\begin{eqnarray}
R_{ab}= {1\over{2}}\,\phi_{|a}\,\phi_{|b} + 2\,T^{EM}_{ab}
\ ,
\label{ricci}
\end{eqnarray}
and the electromagnetic energy-momentum tensor is
\begin{eqnarray}
T^{EM}_{ij} = e^{-a\,\phi}\,\left[ F_{ik}\,F^k_j-{1\over{4}}\,g_{ij}\,
F^2\right]
\equiv
e^{-a\,\phi}\,F\,F\,g\,\left(1-{1\over{4}}\,g\,g\right)
\ ,
\end{eqnarray}
where, in the far R.H.S. we omit indices for brevity, so that $F$
represents any of the components of the Maxwell field strength and
$g$ (not to be confused with the wave parameter that we have set to 1)
any component of the metric tensor.
\par
For the purpose of further simplifying the expressions, we also observe
that Eq.~(\ref{grav}) above can be formally rewritten as
\begin{eqnarray}
{\cal G}\,G={\cal G}\,R
\ ,
\label{grav1}
\end{eqnarray}
where ${\cal G}$ represents any differential operator with coefficients
which depend on gravitational quantities only, $G$ represents any
gravitational quantity and $R$ here stands for any component of the
Ricci tensor.
On omitting vanishing terms one then finds that at order $(1,2)$
Eq.~(\ref{grav}) can be written
\begin{eqnarray}
k_G\,\left({\cal G}^{(0,0)}\,G^{(1,2)}+{\cal G}^{(0,2)}\,G^{(1,0)}
+{\cal G}^{(1,0)}\,G^{(0,2)}+{\cal G}^{(1,2)}\,G^{(0,0)}\right)
&=&{\cal G}^{(0,0)}\,R^{(1,2)}
+k_G\,{\cal G}^{(1,0)}\,R^{(0,2)}
\ ,
\label{g12}
\end{eqnarray}
where
\begin{eqnarray}
R^{(0,2)}&=&2\,F^{(0,1)}\,F^{(0,1)}\,g^{(0,0)}\,
\left(1-{1\over{4}}\,g^{(0,0)}\,g^{(0,0)}\right)
\nonumber \\
R^{(1,2)}&=&k_d\,\phi_{|a}^{(1,0)}\,\phi_{|b}^{(0,2)}
+4\,k_{EM}\,F^{(0,1)}\,F^{(1,1)}\,g^{(0,0)}\,
\left(1-{1\over{4}}\,g^{(0,0)}\,g^{(0,0)}\right)
\nonumber \\
&&+2\,F^{(0,1)}\,F^{(0,1)}\,\left[
k_G\,g^{(1,0)}\,\left(1-{3\over 4}\,g^{(0,0)}\,g^{(0,0)}\right)
-a\,k_d\,\phi^{(1,0)}\,g^{(0,0)}\right]
\ .
\end{eqnarray}
If we now apply our ansatz, Eq.~(\ref{kkk}), and neglect any terms
proportional to $k_G$, $k_d$ with respect to terms proportional to
$k_{EM}$, we find that the third equation in Eq.~(\ref{grav}) is not
affected, since it does not explicitly depend on electromagnetic waves,
whilst the first and second equations can apparently be reduced to
\begin{eqnarray}
0\simeq {\cal G}^{(0,0)}\,R^{(1,2)}\simeq
4\,k_{EM}\,{\cal G}^{(0,0)}\,F^{(0,1)}\,F^{(1,1)}\,g^{(0,0)}\,
\left(1-{1\over{4}}\,g^{(0,0)}\,g^{(0,0)}\right)\,
\ .
\label{g12c}
\end{eqnarray}
These would be undesired constraints for the solutions of Maxwell's
equations at order $(1,1)$ which we have already found in \cite{kndw}.
However, we recall here that those (particular) solutions of the
inhomogeneous equations which we denoted by $\phi_i^{(1,1)}$ in
Section IV of \cite{kndw} are proportional to $k_d/k_{EM}$.
Therefore $F^{(1,1)}\sim k_d/k_{EM}$ and, on substituting into the
R.H.S. of Eq.~(\ref{g12c}), one obtains expressions which are indeed
proportional to $k_d$.
Thus it is not consistent to neglect terms proportional to $k_G$,
$k_d$ in Eq.~(\ref{g12}), since there is no term proportional to
$k_{EM}$ with respect to which they are small and none of the above
gravitational equations is affected by the approximation in
Eq.~(\ref{kkk}).
They are and remain three independent equations for the gravitational
quantities $G^{(1,2)}$.
\par
Of course, Eq.~(\ref{g12}) is so involved that obtaining an analytic
solution
appears to be unlikely.
\subsection{Dilaton equation}
\label{dila_w}
The equation for the dilaton field in the NP tetrad components is
\begin{eqnarray}
&&\left[\hat D\,\hat\Delta+\hat\Delta\,\hat D
-\hat\delta\,\hat\delta^*-\hat\delta^*\,\hat\delta
+(\epsilon+\epsilon^*-\tilde\rho-\tilde\rho^*)\,\hat\Delta
+(\mu+\mu^*-\gamma-\gamma^*)\,\hat D\right.
\nonumber \\
&&\left.\phantom{[}
+(\tau-\pi^*+\tilde\alpha^*-\beta)\,\hat\delta^*
+(\tau^*-\pi+\tilde\alpha-\beta^*)\,\hat\delta\right]\phi=
-a\,e^{-a\,\phi}\,F^2
\ ,
\end{eqnarray}
and can be rewritten in analogy with Eq.~(\ref{grav1}) as
\begin{eqnarray}
{\cal G}\,\phi=-a\,e^{-a\,\phi}\,F\,F\,g\,g
\ .
\label{dil}
\end{eqnarray}
At order $(1,2)$ and neglecting terms which vanish identically,
one then has
\begin{eqnarray}
k_d\,{\cal G}^{(0,0)}\,\phi^{(1,2)}
+k_d\,{\cal G}^{(0,2)}\,\phi^{(1,0)}
+k_G\,{\cal G}^{(1,0)}\,\phi^{(0,2)}
&=&-2\,a\,k_{EM}\,F^{(1,1)}\,F^{(0,1)}\,g^{(0,0)}\,g^{(0,0)}
\nonumber \\
&&-2\,a\,k_G\,F^{(0,1)}\,F^{(0,1)}\,g^{(1,0)}\,g^{(0,0)}
\nonumber \\
&&+a^2\,k_d\,\phi^{(1,0)}\,F^{(0,1)}\,F^{(0,1)}\,g^{(0,0)}
\ .
\end{eqnarray}
Again, we recall that $F^{(1,1)}\sim k_d/k_{EM}$, so that no term
proportional to $k_{EM}$ appears, and therefore the dilaton equation
at order $(1,2)$ does not lead to an undesired constraint for lower
order quantities.
\subsection{Maxwell equations}
\label{max_w}
The equations for the Maxwell fields in the NP formalism
are given by
\begin{eqnarray}
&&(\hat D-2\,\tilde\rho)\,\phi_1
-(\hat\delta^*+\pi-2\,\tilde\alpha)\,\phi_0
+\kappa\,\phi_2=J_{1}
\nonumber \\
&&(\hat\delta-2\,\tau)\,\phi_1
-(\hat\Delta+\mu-2\,\gamma)\,\phi_0
+\sigma\,\phi_2=J_{3}
\nonumber \\
&&(\hat D-\tilde\rho+2\,\epsilon)\,\phi_2
-(\hat\delta^*+2\,\pi)\,\phi_1
+\lambda\,\phi_0=J_{4}
\nonumber \\
&&(\hat\delta-\tau+2\,\beta)\,\phi_2
-(\hat\Delta+2\,\mu)\,\phi_1
+\nu\,\phi_0=J_{2}
\ ,
\label{max}
\end{eqnarray}
where the source terms are
\begin{eqnarray}
J_{1}&=&{a\over 2}\,\left[(\phi_1+\phi_1^*)\,\hat D
-\phi_0\,\hat\delta^*-\phi_0^*\,\hat\delta\right]\phi
\nonumber \\
J_{2}&=&{a\over 2}\,\left[(\phi_1+\phi_1^*)\,\hat\Delta
-\phi_2\,\hat\delta-\phi_2^*\,\hat\delta^*\right]\phi
\nonumber \\
J_{3}&=&{a\over 2}\,\left[(\phi_1-\phi_1^*)\,\hat\delta
-\phi_0\,\hat\Delta+\phi_2^*\,\hat D\right]\phi
\nonumber \\
J_{4}&=&{a\over 2}\,\left[(\phi_1-\phi_1^*)\,\hat\delta^*
+\phi_0^*\,\hat\Delta-\phi_2\,\hat D\right]\phi
\ .
\label{curr}
\end{eqnarray}
We can expand to order $(1,2)$ and make the same approximation given in
Eq.~(\ref{kkk}).
However, it is clear that now there are terms truly proportional to
$k_{EM}$, namely $\phi_i^{(1,0)}$ and $\phi_i^{(1,2)}$, which will
remain even after considering $k_{EM}\,\phi^{(1,1)}\sim k_d$
as found in \cite{kndw}.
We are then allowed to neglect all terms proportional to $k_G$, $k_d$
in Eqs.~(\ref{max}) and (\ref{curr}) above, and the four Maxwell
equations
at order $(1,2)$ then read
\begin{eqnarray}
&&
(\hat D-2\,\tilde\rho)^{(0,0)}\,\phi_1^{(1,2)}
-(\hat\delta^*+\pi-2\,\tilde\alpha)^{(0,0)}\,\phi_0^{(1,2)}
=J_{1}^{(1,2)}
\nonumber \\
&&
(\hat\delta-2\,\tau)^{(0,0)}\,\phi_1^{(1,2)}
-(\hat\Delta+\mu-2\,\gamma)^{(0,0)}\,\phi_0^{(1,2)}
=J_{3}^{(1,2)}
\nonumber \\
&&
(\hat D-\tilde\rho)^{(0,0)}\,\phi_2^{(1,2)}
-(\hat\delta^*+2\,\pi)^{(0,0)}\,\phi_1^{(1,2)}
=J_{4}^{(1,2)}
\nonumber \\
&&
(\hat\delta-\tau+2\,\beta)^{(0,0)}\,\phi_2^{(1,2)}
-(\hat\Delta+2\,\mu)^{(0,0)}\,\phi_1^{(1,2)}
=J_{2}^{(1,2)}
\ .
\label{max12}
\end{eqnarray}
The currents on the R.H.S.s are given by
\begin{eqnarray}
J_{1}^{(1,2)} &=&
-i\,\left({K\over\Delta}\right)^{(0,2)}\,\phi_1^{(1,0)}
+{a\over 2}\,\left[(\phi_1+\phi_1^*)^{(1,0)}\,\partial_r
-{1\over\sqrt{2}}\,\left({\phi_0\over\bar\rho^*}
+{\phi_0^*\over\bar\rho}\right)^{(1,0)}\,\partial_\theta
\right]\,\phi^{(0,2)}
\nonumber \\
J_{2}^{(1,2)} &=&
2\,\mu^{(0,2)}\,\phi_1^{(1,0)}
-{a\over 2}\,\left[{\Delta_0\over 2\,\rho^2}\,(\phi_1+\phi_1^*)^{(1,0)}\,
\partial_r
+{1\over\sqrt{2}}\,\left({\phi_2^*\over\bar\rho^*}
+{\phi_2\over\bar\rho}\right)^{(1,0)}\,\partial_\theta
\right]\,\phi^{(0,2)}
\nonumber \\
J_{3}^{(1,2)} &=&
-\mu^{(0,2)}\,\phi_0^{(1,0)}
+{a\over 2}\,\left[{1\over\sqrt{2}\,\bar\rho}\,
(\phi_1-\phi_1^*)^{(1,0)}\,\partial_\theta
+\left({\Delta_0\over2\,\rho^2}\,\phi_0+\phi_2^*\right)^{(1,0)}\,
\partial_r\right]\,\phi^{(0,2)}
\nonumber \\
J_{4}^{(1,2)} &=&
-i\,\left({K\over\Delta}\right)^{(0,2)}\,\phi_2^{(1,0)}
+{a\over 2}\,\left[{1\over\sqrt{2}\,\bar\rho^*}\,
(\phi_1-\phi_1^*)^{(1,0)}\,\partial_\theta
-\left({\Delta_0\over2\,\rho^2}\,\phi_0^*+\phi_2\right)^{(1,0)}\,
\partial_r
\right]\,\phi^{(0,2)}
\ ,
\label{curr12}
\end{eqnarray}
from which one concludes that the perturbations $\phi_i^{(1,2)}$
couple to the (gradient of the) dilaton background through the free
Maxwell waves $\phi_i^{(1,0)}$.
\section{Asymptotic solutions: scattering from the background}
\label{sol}
The set of equations (\ref{max12}) together with the corresponding
currents is still quite involved.
However, we observe that the electromagnetic waves $\phi_i^{(1,0)}$
which enter the currents in Eq.~(\ref{curr12}) are actually input data,
that is we are free to chose whatever kind of electromagnetic
waves we want to send toward the black hole and then compute the
scattered pattern $\phi_i^{(1,2)}$.
\par
Of course, one wants to consider a physically meaningful model.
For instance, one can think of a double system made of a black hole
and a companion star which is periodically occultated while revolving
around the black hole.
The light of the star would thus periodically pass through the ergoregion
of the black hole where the dilaton background gradient is the strongest.
This would allow for a comparison between the spectrum of the star when
it is in front of the hole (and the dilaton background effects are
negligible) and the spectrum of the star when it is just going behind
the black hole (see Fig.~1).
We also note here that a classical wave scattered by the system is
represented by a superposition of ingoing modes coming from far away
until it reaches the center of the system (the black hole), and then it
switches to a superposition of outgoing modes moving away from the
black hole.
\begin{figure}
\leavevmode
\centerline{\epsfysize=200pt\epsfbox{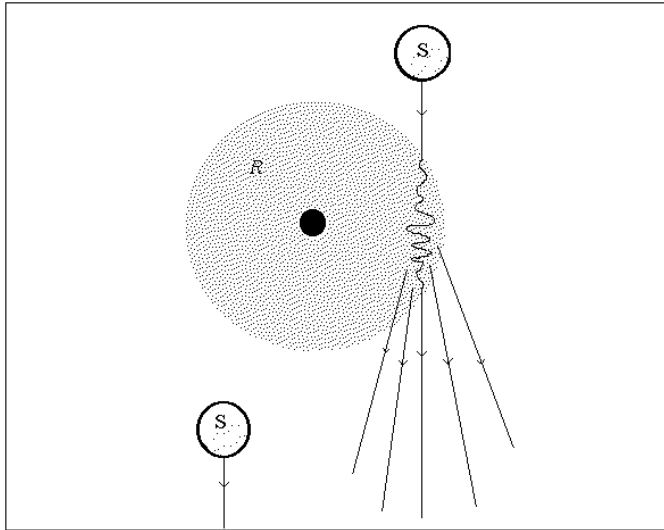}}
\caption{A black hole binary system with the stellar companion shown at
two different positions in its orbit about the black hole.  When the star
is occulting as seen from the earth, electromagnetic radiation in the
earth's direction will be scattered by the tensor and scalar components
of the gravitational field.}
\label{fig1}
\end{figure}

\par
Asymptotically ($r\to\infty$) the major contribution to the
electromagnetic wave field is given by
\begin{eqnarray}
\phi_2^{(1,0)}\simeq C_2^\pm\,{e^{\pm i\,\bar\omega\,r}\over r}
\equiv F_2^\pm
\ ,
\end{eqnarray}
with a plus sign for ingoing modes and a minus sign for outgoing modes
and $C_2^\pm=C_2^\pm(\theta)$ contains all the angular dependence in
$\theta$.
Therefore we can assume that $\phi_0^{(1,0)}=\phi_1^{(1,0)}\equiv 0$
(the so called {\em phantom gauge}, see Ref.~\cite{chandra})
in Eq.~(\ref{curr12}).
This reduces the currents to
\begin{eqnarray}
J_{1}^{(1,2)} &=&0
\nonumber \\
J_{2}^{(1,2)} &=&
-{a\over 2\,\sqrt{2}}\,\left({F_2^*\over\bar\rho^*}
+{F_2\over\bar\rho}\right)\,\partial_\theta\phi
\nonumber \\
J_{3}^{(1,2)} &=&
{a\over 2}\,F_2^*\,\partial_r\phi
\nonumber \\
J_{4}^{(1,2)} &=&i\,{K\over\Delta_0^2}\,F_2
-{a\over 2}\,F_2\,\partial_r\phi
\ ,
\label{curr122}
\end{eqnarray}
where
\begin{eqnarray}
\phi\equiv\phi^{(0,2)}=
-{a\over M}\,{r\over\rho^2}
\ .
\end{eqnarray}
\par
We can obtain an equation for $\phi_2^{(1,2)}$ alone from the third
and fourth equations in Eq.~(\ref{max12}).
Upon defining as usual $W_2\equiv 2\,(\bar\rho^*)^2\,\phi_2^{(1,2)}$
and $W_1\equiv\sqrt{2}\,\bar\rho^*\,\phi_1^{(1,2)}$ we obtain
\begin{eqnarray}
&&
\left({\cal D}_0-{1\over\bar\rho^*}\right)\,W_2
-\left({\cal L}_0+{i\,\alpha\,\sin\theta\over\bar\rho^*}\right)\,W_1
=2\,(\bar\rho^*)^2\,F_2\,\left(i\,{K\over\Delta_0^2}-
{a\over 2}\,\partial_r\phi\right)
\nonumber \\
&&
\left({\cal L}_1^\dagger-{i\,\alpha\,\sin\theta\over\bar\rho^*}\right)\,
W_2
-\Delta_0\,\left({\cal D}_0^\dagger+{1\over\bar\rho^*}\right)\,W_1
=-a\,\bar\rho^*\,(\bar\rho^*\,F_2+\bar\rho\,F_2^*)\,
\partial_\theta\phi
\ .
\end{eqnarray}
Finally, upon using the commutation relation
\begin{eqnarray}
\left[\Delta_0\,\left({\cal D}_0^\dagger+{1\over\bar\rho^*}\right),
\left({\cal L}_0+{i\,\alpha\,\sin\theta\over\bar\rho^*}\right)
\right]=0
\ ,
\end{eqnarray}
one can eliminate $W_1$ and gets
\begin{eqnarray}
\left[\Delta_0\,{\cal D}_0^\dagger\,{\cal D}_0
+{\cal L}_0\,{\cal L}_1^\dagger+2\,i\,\bar\omega\bar\rho\right]\,W_2
=T_2
\ ,
\label{W2}
\end{eqnarray}
where the source term is
\begin{eqnarray}
T_2&\equiv&
2\,\Delta_0\,\left({\cal D}_0^\dagger+{1\over\bar\rho^*}\right)\,
\left[(\bar\rho^*)^2\,F_2\,\left(i\,{K\over\Delta_0^2}-
{a\over 2}\,\partial_r\phi\right)\right]
\nonumber \\
&&
-a\,\left({\cal L}_0+{i\,\alpha\,\sin\theta\over\bar\rho^*}\right)\,
\left[\bar\rho^*\,(\bar\rho^*\,F_2+\bar\rho\,F_2^*)\,
\partial_\theta\phi
\right]
\ .
\end{eqnarray}
\par
To leading order in $1/r$, the current is thus
\begin{eqnarray}
T_2^\pm=2\,\bar\omega\,C_2^\pm\,(2\,M\,\bar\omega+i\,a^2)\,
\left\{\begin{array}{l}
\,e^{+i\,\bar\omega\,r}
\\
{r\over M}\,e^{-i\,\bar\omega\,r}
\ .
\end{array}\right.
\end{eqnarray}
In the expression above the contribution from the dilaton background
is the one proportional to $a^2$, the remaining terms being purely KN.
We can then write
\begin{eqnarray}
T_2=T_{KN}+T_a
\ ,
\end{eqnarray}
and separate the contributions for the two different sources.
The L.H.S. of Eq.~(\ref{W2}) can be expanded in powers of $1/r$ as well
upon assuming
\begin{eqnarray}
W_2\simeq A^\pm\,Z^\pm\,{e^{\pm i\,\bar\omega\,r}\over r^{n_\pm}}
\ ,
\end{eqnarray}
where $Z^\pm=Z^\pm(\theta)$.
One then finds
\begin{eqnarray}
A^\pm\,Z^\pm\,{e^{\pm i\,\bar\omega\,r}\over r^{n_\pm-1}}\,
2\,\bar\omega\,\left[2\,M\,\bar\omega+i\,(1\mp n_\pm)\right]
= T_2^{\pm} \ ,
\end{eqnarray}
from which it follows that $Z^\pm=C_2^\pm$,
\begin{eqnarray}
\left\{\begin{array}{l}
n_+=1 \\
A^+_{KN}=1 \\
A^+_a=i\,{a^2\over2\,M\,\bar\omega}
\ ,
\end{array}\right.
\end{eqnarray}
and
\begin{eqnarray}
\left\{\begin{array}{l}
n_-=0 \\
A^-_{KN}=2\,\bar\omega\,{2\,M\,\bar\omega-i\over1+4\,M^2\,\bar\omega^2}
\\
A^-_a={a^2\over M}\,{1+2\,i\,M\,\bar\omega\over1+4\,M^2\,\bar\omega^2}
\ .
\end{array}\right.
\end{eqnarray}
We can conclude that the amplitudes of the electromagnetic waves
scattered
by the dilaton background become comparable to or greater than the
intensity of the
waves scattered by the KN background when $A_a^-\ge A_{KN}^-$,
that is when (we restore the fundamental constants)
\begin{eqnarray}
\bar\omega\le\bar\omega_c\sim {a^2\,c^3\over G\,M}
\ .
\end{eqnarray}
In string theory $a=1$ and $\bar\omega_c\sim 10^{35}$ Hz$/$kg.
For example, for a solar mass black hole, the critical frequency would
be about 100 kHz.
\par
We observe that $\phi_2^{(1,2)}\sim 1/r^2$ is subleading
with respect to $\phi_2^{(1,0)}\sim 1/r$.
However, we can apply the same argument that we have formulated in
\cite{kndw}, Section V, and conclude that the scattered waves
$\phi_2^{(1,2)}$ are negligible with respect to free waves
$\phi_2^{(1,0)}$ only when the scattering process occurs at large $r$,
which is quite sensible, since strong effects from
the dilaton background are not expected far away from the horizon.
Therefore, adjusting the above mentioned argument to the present context,
we claim that testable electromagnetic waves could be produced by the
scattering of free Maxwell waves in a region (denoted by ${\cal R}$ in
Fig.~1 and Ref.~\cite{kndw}) just outside the horizon where the
gradient of the dilaton background is the strongest.
Outside of ${\cal R}$ the scattered waves then propagate as free waves,
thus one obtains
\begin{eqnarray}
|\phi_2^{(1,2)}(r>r_d)|\simeq|\phi_2^{(1,2)}(r_d)|\,{r_d\over r}
\ ,
\end{eqnarray}
where $r_d$ is the typical outer radial coordinate of the region
${\cal R}$.
\par
The next step would now be to superpose a suitable selection of incoming
modes $\phi_2^{(1,0)}$  to more realisitically model the electromagnetic
radiation coming
from the stellar companion of the black hole.
Then, the corresponding superposition of excited modes $\phi_2^{(1,2)}$
would give us the scattered pattern and its time dependence due to the
relative positions of the source and the black hole with respect to
the observer.
However, we feel that sensible results are out of reach of the analytical
methods which are all we want to consider in the present work.
\section{Geometrical optics effects}
\label{def}
In the previous sections and in \cite{kndw} we have studied the wave
equations on the KND background.
However, one might expect to obtain measurable effects on the propagation
of light in such a background even at the level of geometrical optics
\cite{mtw}.
\par
To start with, we consider a simpler case, the (exact) RND solution
\cite{rnd},
\begin{eqnarray}
ds^2=-e^{2\,\Phi}\,dt^2+e^{2\,\Lambda}\,dr^2
+R^2\,\left(d\theta^2+\sin^2\theta\,d\varphi^2\right)
\ ,
\label{grnd}
\end{eqnarray}
with
\begin{eqnarray}
&&e^{2\,\Phi}=e^{-2\,\Lambda}
=\left(1-{r_+\over r}\right)\,
\left(1-{r_-\over r}\right)^{1-a^2\over 1+a^2}
\nonumber \\
&&R^2=r^2\,\left(1-{r_-\over r}\right)^{2\,a^2\over 1+a^2}
\nonumber \\
&&r_+=M\,\left[1+\sqrt{1-(1-a^2)\,{Q^2\over M^2}}\; \right]
\nonumber \\
&&r_-=(1+a^2)\,{Q^2\over M}\,
\left[1+\sqrt{1-(1-a^2)\,{Q^2\over M^2}}\; \right]^{-1}
\ ,
\end{eqnarray}
and compute the deflection angle $\Delta\varphi_1$ of a null ray coming
from infinity with an impact parameter $D_1=L_1/E_1\equiv D$
($L_1$ is its angular momentum, $E_1$ its energy) and approaching the
outer horizon $r_+$.
Fromthe equations of motion for null geodesics one obtains an expression
for $d\varphi/dr$ (see \cite{chandra} for the details).
On integrating the latter from $r=+\infty$ to $r_1$,
the minimum radial position from the center of the hole reached by the ray,
one finds
\begin{eqnarray}
\Delta\varphi_1=2\,\left(\varphi(r_1)-\varphi(\infty)\right)-\pi
\ .
\end{eqnarray}
For the metric in Eq.~(\ref{grnd}) the above expression becomes
\begin{eqnarray}
\Delta\varphi_1\simeq
{2\,M\over r_1}\,\left[1+{4\,a^2-3\over 1+a^2}\,{Q^2\over 4\,M^2}\right]
\ ,
\label{dphi}
\end{eqnarray}
where we have assumed $r_+\gg r_-$ and $Q/M\ll 1$.
Eq.~(\ref{dphi}) already contains a dilatonic contribution and allows a
comparison of the pure Reissner-Nordstr\"om case ($a=0$) with the
Schwarzschild case ($Q=0$).
\par
The radius $r_1$ is one of the zeros of $dr/d\varphi$ which in turn
are the solutions of \cite{chandra}
\begin{eqnarray}
r_1^3-D_1^2\,(r_1-r_+)=0
\ .
\label{r}
\end{eqnarray}
Of course one requires $r_1>r_+$ for an unbound null ray that must escape
to infinity.
\par
Let us now consider a second fiducial null ray starting far away from
the hole which lies in the same plane and points in the same direction
as the first one but has a slightly different impact parameter
$D_2=J_2/E_2=D+\delta$, where $\delta \ll D$.
The two rays, rotated around the axis passing through the center of the
hole and asymptotically parallel to both of them, define an annulus of
which we just consider a small portion whose area is
\begin{eqnarray}
A_{(-\infty)}\sim \Theta\,(D_2^2-D_1^2)\simeq
2\,D\,\delta\,\Theta
\ ,
\end{eqnarray}
where $\Theta$ is the relative angle between the two rays with respect
to the axis passing through the center of the hole
(we assume $\Theta$ to be small).
\par
Each ray will then be deflected by the gravitational field of the hole
according to Eq.~(\ref{dphi}).
Since one has $r_1\not=r_2$, after having been scattered, the two
fiducial rays will move at a relative angle
\begin{eqnarray}
\Delta_{12} \equiv \Delta\varphi_2-\Delta\varphi_1
=2\,M\,\left[1+{4\,a^2-3\over 1+a^2}\,{Q^2\over 4\,M^2}\right]\,
\left({1\over r_2}-{1\over r_1}\right)
\ .
\label{delta}
\end{eqnarray}
Therefore the area that they define will change according to
\begin{eqnarray}
A_{(r)}\sim r^2\,\Theta\,|\Delta_{12}|
\ .
\end{eqnarray}
\par
Although Eq.~(\ref{r}) is exactly soluable, the solutions are complicated,
thus we will just consider the following approximations.
Since we are interested in rays which travel close to the horizon,
we assume $r_i=r_++x_i$, with $x_i\ll r_+$ and, from Eq.~(\ref{r}) we get
\begin{eqnarray}
D_i^2\simeq {r_+^3\over x_i}\gg r_+^2
\ .
\label{D}
\end{eqnarray}
On solving the above equation for $x_i$ and substituting into
Eq.~(\ref{delta}) one finally obtains
\begin{eqnarray}
\Delta_{12}\simeq
{16\,M^2\,\delta\over D^3}\,
\left[1+{2\,a^4+4\,a^2-5\over 1+a^2}\,{Q^2\over 8\,M^2}\right]
\ ,
\end{eqnarray}
where the approximation $D_i\gg r_+,\delta$ has been used, but $\delta
\sim r_+$ is allowed.
\par
The rate of decrease of the intensity $I$ can then be computed for any
massless linear wave moving ``between the two rays'',
\begin{eqnarray}
{I_{(r)}\over I_{(-\infty)}}={A_{(-\infty)}\over A_{(r)}}
\sim {D^4\over M^2}\,
\left[1-{2\,a^4+4\,a^2-5\over 1+a^2}\,{Q^2\over 8\,M^2}\right]\,
{1\over r^2}
\ .
\end{eqnarray}
\par
The above result depends on the form of the metric only, and one might
think of repeating the same computation for the KND case.
However, as we pointed out in \cite{kndw}, up to the order at which we
have
computed it, the KND metric is just the KN metric.
Therefore, no effect on the null rays can be expected from the dilaton
background at any order below $Q^4/M^4$ when the black hole is rotating
\cite{es}.
\par
A final remark is due to clarify the difference between  the approach
shown in this section and the previous one.
The solutions to the wave equation that we have found in section~\ref{sol}
display the full wave nature of the electromagnetic radiation and they
also depend on the direct coupling between the dilaton field and the
Maxwell field as formulated in the action (\ref{action}).
However, the results in the present section rely on the approximation of
geometrical optics, thus neglecting the specific nature of the null rays
which includes both the wave character and the coupling to the dilaton.
Therefore it is not surprising that, in the geometrical optics picture
no affect on the Maxwell rays is found for the KND geometry, while in
the wave picture and at the same order of approximation in the
charge-to-mass expansion the dilaton does affect electromagnetic waves.
\par
By comparing the two approaches, one can conclude that the presence of a
non-trivial background dilaton in KND (at the order included in our
computation) affects only the intensity of the electromagnetic radiation
and not its eikonal paths.
This means that the calculation of gravitational lensing effects for
comparison with measurements as tests
for the existence of non-trivial dilaton fields must be carried to higher
order in the expansion,  and this can probably be done only numerically.
\section{Conclusions}
We have shown that to the order at which we are working in perturbation
theory the neglect of the  gravitational and dilatonic waves with respect
to the electromagnetic waves is internally consistent. 
This assumption has allowed us to obtain asymptotic
expressions for ingoing and outgoing modes in the phantom gauge and the
dilatonic contributions to these modes. 
Our analysis shows that electromagnetic waves scattered by the dilaton
background have a critical frequency of approximately 0.1~MHz. 
This is in a frequency range which is not currently being studied by
astrophysicists, because most stars emit relatively small amounts of
energy in this range. 
On the other hand, if a detector were to be constructed to observe
black hole binaries in this frequency range, the signal should be
relatively clean, and any enhancement at this frequency in the intensity
of the radiation from an occulting star would be a clear indication of
the presence of a scalar component of gravity.
\par
At any given frequency a decrease in the intensity of the radiation should
be expected for an occulting star due to the scattering of the light by the
gravitational field. 
However an exact expression for the intensity will have to wait until we
are able to realistically model the electromagnetic radiation from the
stellar companion of a black hole.
This involves the superposition of the incoming Maxwell scalar modes,
and this will probably entail a numerical investigation of the variation
of the intensity of the star as a function of its orbital position.
\par
In the geometrical approximation for a charged, non-rotating
(Reissner-Nordstr\"om) dilaton black hole the integrated intensity
decreases during occultation if $2\,a^4+4\,a^2-5 > 0$. 
In string theory $a=1$, and  a decrease in the intensity of light from
the occulting star as compared to its intensity when it eclipses
the black hole would be evidence for a scalar component of gravity
as predicted by string theory.
\acknowledgments
We wish to thank Y. Leblanc for his contributions to the early stages
of this work. 
This work was supported in part by the U.S. Department
of Energy under Grant No. DE-FG02-96ER40967.

\begin{references}
\bibitem{es}
This conclusion seems to hold also for the metric in the so called
{\em string frame} which is obtained from the one we have been using
so far by a conformal transformation.
Work in preparation.
\end{references}
\end{document}